\documentclass[%
 showpacs,
 amsmath,amssymb,
]{revtex4-1}

\usepackage{graphicx}
\usepackage{dcolumn}
\usepackage{bm}

\usepackage{CJK}

\usepackage{color}
\usepackage{amssymb}
\usepackage{amsfonts}

\usepackage[colorlinks,urlcolor=blue,linkcolor=blue,citecolor=blue,anchorcolor=blue]{hyperref}

\begin{document}

\preprint{APS/123-QED}

\title{Propagation dynamics of a light beam in fractional Schr\"odinger equation}

\author{Yiqi Zhang$^1$}
\email{zhangyiqi@mail.xjtu.edu.cn}
\author{Xing Liu$^1$}
\author{Milivoj R. Beli\'c$^{2}$}
\author{Weiping Zhong$^3$}
\author{Yanpeng Zhang$^{1}$}
\email{ypzhang@mail.xjtu.edu.cn}
\author{Min Xiao$^{4,5}$}
\affiliation{%
 $^1$Key Laboratory for Physical Electronics and Devices of the Ministry of Education \& Shaanxi Key Lab of Information Photonic Technique,
Xi'an Jiaotong University, Xi'an 710049, China \\
$^2$Science Program, Texas A\&M University at Qatar, P.O. Box 23874 Doha, Qatar \\
$^3$Department of Electronic and Information Engineering, Shunde Polytechnic, Shunde 528300, China \\
$^4$Department of Physics, University of Arkansas, Fayetteville, Arkansas, 72701, USA \\
$^5$National Laboratory of Solid State Microstructures and School of Physics, Nanjing University, Nanjing 210093, China
}%

\date{\today}

\begin{abstract}
  \noindent
  Dynamics of wavepackets in fractional Schr\"odinger equation is still an open problem.
  The difficulty stems from the fact that the fractional Laplacian derivative is essentially a nonlocal operator.
  We investigate analytically and numerically the propagation of optical beams in fractional Schr\"odinger equation with a harmonic potential.
  We find that the propagation of
  one- and two-dimensional (1D, 2D) input chirped Gaussian beams is not harmonic.
  In 1D, the beam propagates along a zigzag
  trajectory in the real space, which corresponds to a
  modulated anharmonic oscillation in the momentum space.
  In 2D, the input Gaussian beam
  evolves into a breathing ring structure in both real and momentum spaces, which forms a filamented funnel-like aperiodic structure.
  The beams remain localized in propagation, but  with increasing distance display increasingly irregular behavior,
  unless both the linear chirp and the transverse displacement of the incident beam are zero.
\end{abstract}

\maketitle

%

Fractional quantum mechanics is a promising extension of quantum
mechanics that received great attention since its introduction by
Laskin
\cite{laskin.pla.268.298.2000,*laskin.pre.62.3135.2000,*laskin.pre.66.056108.2002}.
It is based on the fractional Schr\"odinger equation (FSE) -- a
generalization of the Schr\"odinger equation (SE) which involves fractional spatial
derivatives instead of the usual ones
 \cite{li.jmp.46.103514.2005,dong.jmp.48.072105.2007,kowalski.pra.81.012118.2010,oliveira.jpa.44.185303.2011,lorinczi.jde.253.2846.2012,luchko.jmp.54.012111.2013,*stickler.pre.88.012120.2013,*zaba.jmp.55.092103.2014,*tare.pa.407.43.2014}.
Most of current research is focused on mathematical issues and the
steady behavior of wavepackets in simple potentials.  However, even
the relativistic massless harmonic oscillator
\cite{li.jmp.46.103514.2005,kowalski.pra.81.012118.2010,lorinczi.jde.253.2846.2012}
associated with FSE turned out not to be so simple.
The difficulty comes from the fact that the fractional Laplacian derivative, which sits in the FSE instead of the ordinary Laplacian, is inherently a nonlocal operator.
Here, we address the dynamics of wavepackets in harmonic potential, but from
a different viewpoint. We base analysis on the equivalence of
paraxial wave equation and the SE, which can be extended to the FSE, and
focus on the differences in real and frequency spaces. Recently, an
optical realization of FSE was advanced by Longhi
\cite{longhi.ol.36.2883.2015}, based on aspherical optical cavities.
He obtained the eigenmodes of a massless harmonic oscillator -- the
dual Airy functions \cite{lorinczi.jde.253.2846.2012,longhi.ol.36.2883.2015}.
Our approach is different; we focus on the dynamics of optical beams in
FSE with harmonic potential and not on the eigenvalue problem. Thus, a realistic input beam is launched into the system and
its dynamics followed, with an emphasis on the differences from the
usual quantum harmonic oscillator.

Similar to the standard SE, the potentials introduced into the FSE
\cite{herrmann.book} can vary widely
\cite{rechtsman.nature.496.196.2013,*zhang.lpr.9.331.2015}.
Among various potentials, the harmonic potential -- broadly used in
quantum mechanics, most notably in Bose-Einstein condensates
\cite{dalfovo.rmp.71.463.1999,*staliunas.prl.89.210406.2002,*liang.pr.94.050402.2005,*bagnato.rrp.28.251.2015},
laser-plasma physics \cite{berge.pp.4.1227.1997,*berge.pr.303.259.1998,*berge.rpp.70.1633.2007},
ultracold atoms \cite{martiyanov.prl.105.030404.2010,*windpassinger.rpp.76.086401.2013,*collura.prl.110.245301.2013},
ion-laser interactions \cite{moyacessa.pr.513.229.2012}, and optical
lattices \cite{kartashov.rmp.83.247.2011}, is probably the most
useful of all. In the linear case, naturally, beams exhibit perfect
harmonic oscillation during propagation in both real and inverse
spaces, because from a mathematical point of view the guiding
equations remain the same in both spaces. However, the dynamics in
the harmonic potential in FSE is different. To the best of our
knowledge, it has not been treated before.

In this Letter, we investigate the dynamics of waves in the FSE with
harmonic potential. We supply an analytical method to study such
dynamics and make comparison with numerical simulation. We discover
that the beams follow zigzag and funnel-like paths in the real space
in one and two dimensions (1D, 2D), which after prolonged
propagation become  irregular. Based on the methods discussed in
Ref. \cite{longhi.ol.36.2883.2015}, results reported in this Letter
can easily be experimentally verified, e.g., they directly apply to
the fractional gradient refractive index (GRIN) media.


The FSE is written as
\begin{equation}\label{58eq1}
  i\frac{\partial \psi}{\partial z} = \left[ \frac{1}{2} \left(-\frac{\partial^2}{\partial x^2}\right)^{\alpha/2} + V(x) \right] \psi,
\end{equation}
where $\alpha$ is the L\'{e}vy index ($1<\alpha\le2$) and $V(x) =
\beta^2 x^2/2$ is the external harmonic potential, with $\beta$
being an arbitrary scaling parameter
\cite{zhang.oe.23.10467.2015,*zhang.ol.40.3786.2015}. Variables $x$
and $z$ are the transverse coordinate and the propagation distance,
scaled by some characteristic transverse width and the corresponding
Rayleigh range \cite{zhang.ol.38.4585.2013,*zhang.oe.22.7160.2014}.
When $\alpha=2$, one recovers the usual SE. Here, we treat the
opposite limiting case $\alpha=1$, as the most interesting
\cite{kowalski.pra.81.012118.2010,lorinczi.jde.253.2846.2012,longhi.ol.36.2883.2015}.
We exploit the fact that in the momentum space (or the $k$ space),
Eq. (\ref{58eq1})  transforms into
\begin{equation}\label{58eq2}
  i\frac{\partial \hat{\psi}}{\partial \xi} + \left( f |k|  + \frac{1}{2} \frac{\partial ^2}{\partial k^2} \right) \hat{\psi} = 0,
\end{equation}
where $\hat{\psi}$ is the Fourier transform of $\psi$, $k$ is the spatial frequency, $\xi=\beta^2z$, and $f=-1/(2\beta^2)$ is a constant.
Clearly, Eq. (\ref{58eq2}) is a normal Sch\"odinger equation with symmetric linear potential,
which brings a \textit{d\'{e}j\`{a} vu} feeling \cite{bernardini.epj.16.58.1995,*gori.ejp.20.477.1999,*liu.ol.36.1164.2011,*kovalev.josaa.31.914.2014}.
If one considers $k>0$ and $k<0$ cases separately,
the solution of Eq. (\ref{58eq2}) can be written as
\begin{align}\label{58eq3}
  \hat{\psi}(k,\xi) = & \sqrt{\frac{1}{2\pi i \xi}} \, \exp\left[ i\left(\pm fk\xi-\frac{f^2\xi^3}{6}\right) \right] \int_{-\infty}^{+\infty}d\kappa 
   \times {\hat\psi}(\kappa,0) \,\exp\left[ \frac{i}{2\xi}\left( k\mp\frac{f
  \xi^2}{2}-\kappa \right)^2 \right],
\end{align}
where $\pm$ corresponds to $k$ positive or negative, respectively. However, the
corresponding solution of Eq. (\ref{58eq1}) cannot be directly
obtained by doing an inverse Fourier transform of Eq. (\ref{58eq3}), due to the non-differentiability of the potential at $k=0$.
That is, it is not so easy to obtain the exact solution of Eq.
(\ref{58eq1}), unless one is only interested in its eigenvalues and
eigenfunctions. Here, we provide an accurate analytical method and
investigate the dynamics of beams with a specific input.

We assume that the input beam is chosen as a common chirped Gaussian
\begin{equation}\label{58eq4}
  \psi(x)=\exp\left[-\sigma(x-x_0)^2\right]\exp(-iCx),
\end{equation}
with $x_0$ being the transverse displacement, $C$ being the linear chirp coefficient, and $\sigma$ controlling the beam width.
The corresponding Fourier transform is
\begin{equation}\label{58eq5}
  {\hat\psi}(k)=\sqrt{\frac{\pi}{\sigma}}\exp\left[-\frac{(k+C)^2}{4\sigma}\right]\exp(-ikx_0),
\end{equation}
which is the initial condition that appears in Eq. (\ref{58eq3}).
Therefore, the corresponding solution can be written as
\begin{align}\label{58eq6}
  \hat{\psi}(k,\xi) = & \sqrt{\frac{\pi}{2A\sigma i \xi}} \, \exp\left[ i\left(\pm fk\xi-\frac{f^2\xi^3}{6}+Cx_0\right)\right] 
   \exp\left[\frac{i}{2\xi}\left(k\mp\frac{f\xi^2}{2}+C\right)^2\right] \exp\left(-\frac{\omega^2}{A}\right) ,
\end{align}
where $A=1/4\sigma-i/2\xi$ and $\omega=k/\xi\mp f\xi/2+C/\xi+x_0$.
Here, {$\pm$ corresponds to the $C\lessgtr0$ case.} From Eq.
(\ref{58eq6}), one can immediately find the trajectory of the beam
in the $k$ space,
\begin{equation}\label{58eq7}
  k=\mp \frac{\beta^2}{4}z^2-x_0\beta^2z-C,
\end{equation}
which is a quadratic function with the symmetry axis
\begin{equation}\label{58eq8}
  z_{\rm sym}=\mp 2x_0,
\end{equation}
and with the sign of $k$ changing at
\begin{equation}\label{58eq9}
  z_0=\mp 2x_0+2\sqrt{x_0^2 \mp \frac{C}{\beta^2}}.
\end{equation}

\begin{figure}[htbp]
  \centering
  \includegraphics[width=0.5\columnwidth]{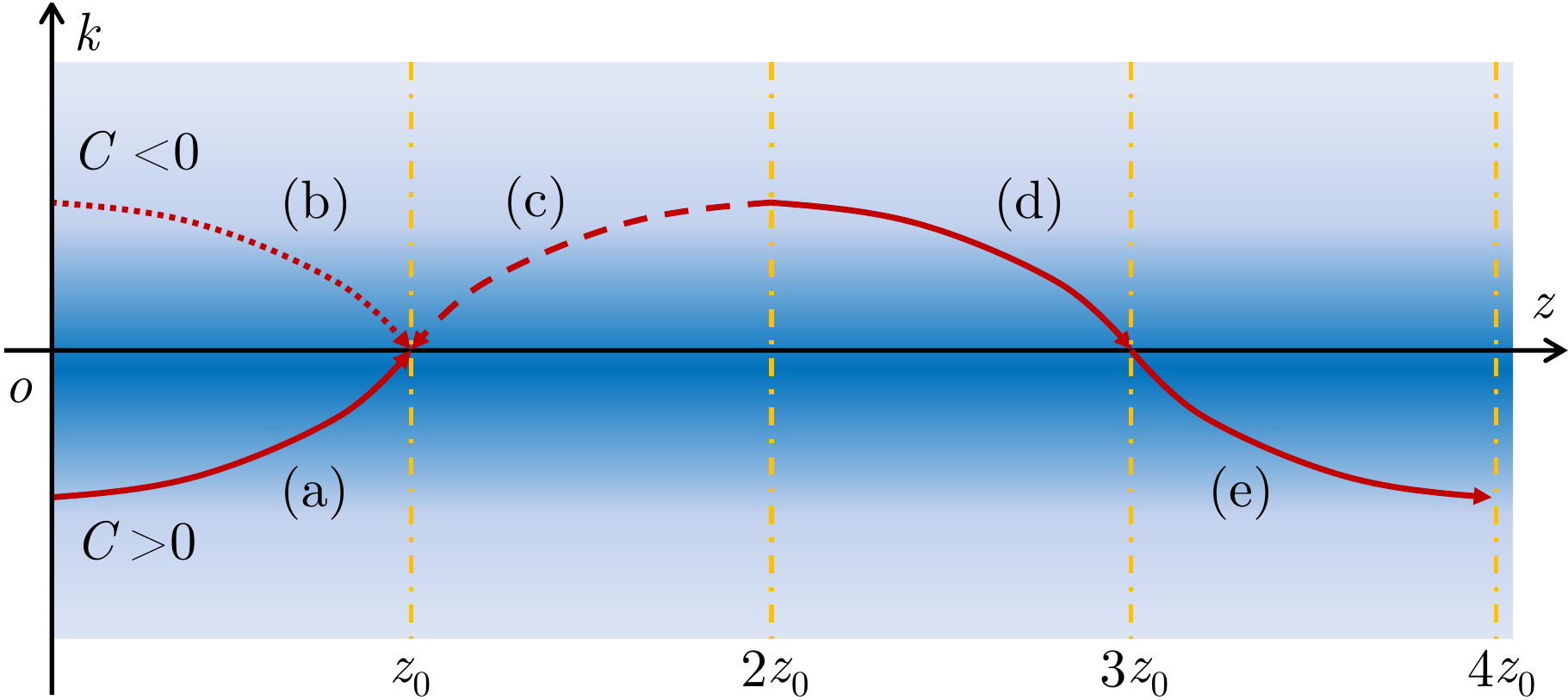}\\
  \caption{(Color online) Schematics of the beam propagation. Here, $x_0=0$.}
  \label{58fig1}
\end{figure}

Now, we analyze the propagation of the beam in such a symmetric linear potential in the inverse space [Eq. (\ref{58eq2})].
We assume $C>0$. The analysis is as follows.\\
(i) In the region $0\le z\le z_0$, the beam propagates as in Fig. \ref{58fig1}(a).\\
(ii) Due to the symmetry of the potential, the propagation of the beam with $-C$ will appear as in Fig. \ref{58fig1}(b).\\
(iii) Continuing with the beam presented in Fig. \ref{58fig1}(b), it will propagate as in Fig. \ref{58fig1}(c),
if it is launched into the medium in the opposite direction.\\
(iv) According to the reciprocal property of the system, the beam from Fig. \ref{58fig1}(a) will propagate along the same trajectory, Fig. \ref{58fig1}(c),
when propagating beyond $z_0$.

Based on the above analysis, the beam with $C$ positive will propagate along the trajectory shown in Figs.
\ref{58fig1}(a), \ref{58fig1}(c), \ref{58fig1}(d) and \ref{58fig1}(e), which forms a whole period of this oscillation.
Clearly, the period is
\begin{equation}\label{58eq10}
  \mathcal{Z}=4(z_0-z_{\rm sym})=8\sqrt{x_0^2 \mp \frac{C}{\beta^2}}.
\end{equation}
Together with Eq. (\ref{58eq9}), one can see that the bigger the chirp and the transverse displacement and the smaller the potential,
the bigger the period $\mathcal{Z}$.

\begin{figure}[htbp]
  \centering
  \includegraphics[width=0.5\columnwidth]{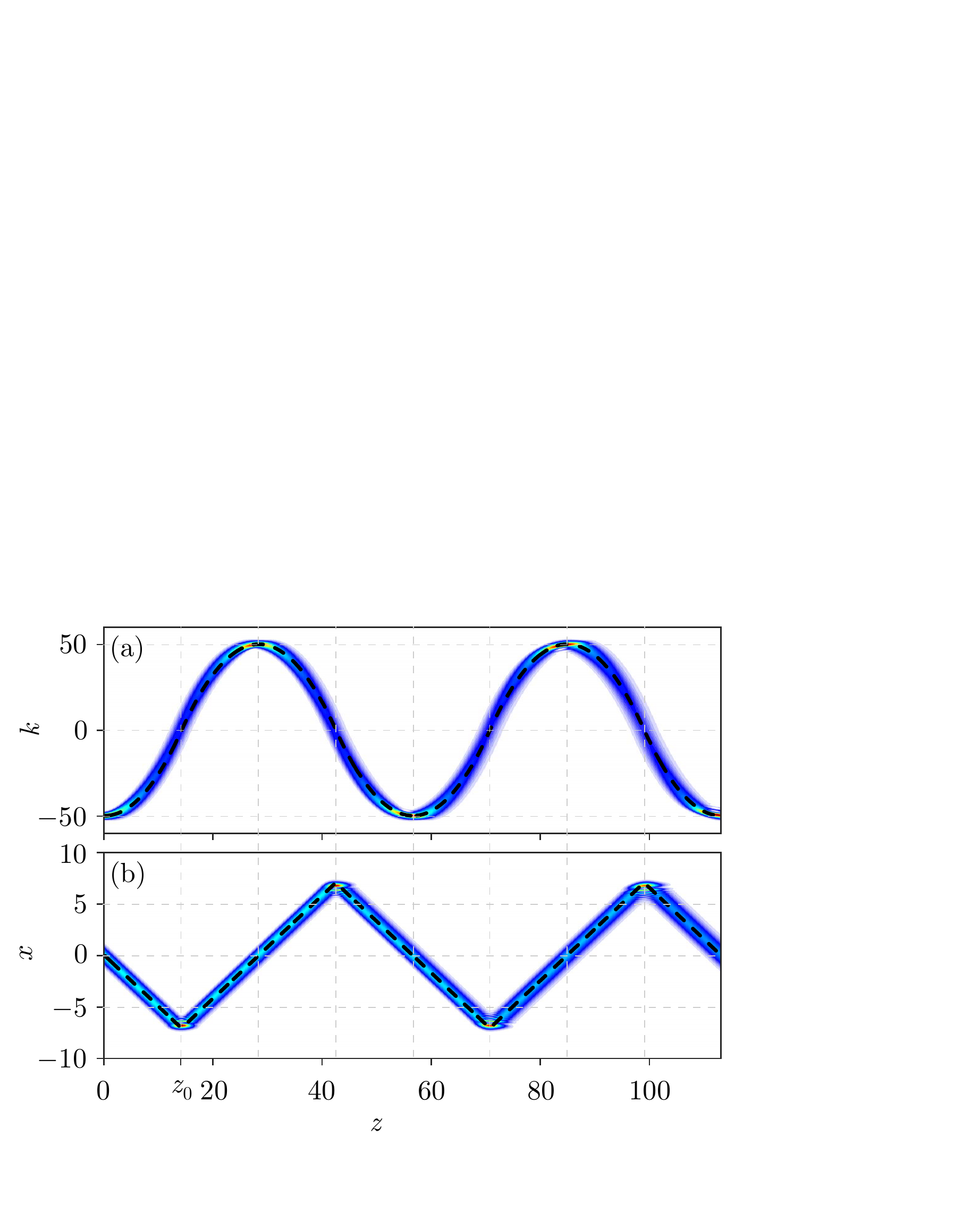}
  \caption{(Color online) Propagation of a chirped Gaussian beam with $\sigma=1$, $x_0=0$, and $C=50$ according to Eq. (\ref{58eq1}) with $\alpha=1$ and $\beta=1$.
  (a) In the $k$ space. (b) In the real space. The dashed curves are analytical trajectories. }
  \label{58fig2}
\end{figure}

In Figs. \ref{58fig2}(a) and \ref{58fig2}(b) we display the
comparison of analytical results and numerical simulations of the
propagation in $k$ and $x$ spaces, with $\beta=1$, $C=50$ and
$x_0=0$. As seen in Fig. \ref{58fig2}(a), the beam exhibits a
modulated oscillation during propagation, in which the numerical and
analytical (the dashed curve) trajectories agree very well.
However, this is not a harmonic oscillation, because the trajectory
is composed of pieces of parabolas. In the $x$ space, as shown in
Fig. \ref{58fig2}(b), the beam propagates along a zigzag trajectory,
which is similar to a total internal reflection trajectory in a
fiber -- in this case, a fractional GRIN fiber.

To analyze the propagation in the real space, it is necessary to perform an inverse Fourier transform of Eq. (\ref{58eq6}),
which is not an easy task. It is found to be
\begin{align}\label{58eq11}
  \psi(x,\xi)=-\sqrt{\frac{1}{8iAA_1\sigma\xi}}
  & \exp\left[i\left(-\frac{f^2\xi^3}{6}+Cx_0\right)\right]
  \exp\left[-\frac{1}{A\xi^2}\left(\mp\frac{f\xi^2}{2}+x_0\xi+C\right)^2\right] \notag \\
\times  & \exp\left[\frac{i}{2\xi}\left(\mp\frac{f\xi^2}{2}+C\right)^2\right]
  \exp\left(\frac{A_2^2}{4A_1}+i\frac{A_2}{2A_1}x-\frac{x^2}{4A_1}\right),
\end{align}
where
$A_1=1/A\xi^2-i/2\xi$
and
$A_2=i(\mp f\xi^2/2+C)/\xi-2(\mp f\xi^2/2+x_0\xi+C)/A\xi^2 - i(\mp f\xi)$.
From Eq. (\ref{58eq11}), one can deduce the trajectory of the beam in the real space,
\begin{equation}\label{58eq12}
  x=\mp\frac{1}{2}z+x_0.
\end{equation}
Corresponding to Figs. \ref{58fig1}(a), \ref{58fig1}(c), \ref{58fig1}(d) and \ref{58fig1}(e),
the beam propagates along the pieces of straight lines.
As a whole, a zigzag trajectory is obtained,
which is shown by the dashed lines in Fig. \ref{58fig2}(b)
and is in full accordance with the numerical simulation.

The physical explanation for the reflection-like behavior is most easily acquired in the $k$ space.
The equivalent refractive index is the biggest along the $k=0$ axis,
so the beam will always be pulled to the axis during propagation, i.e., there is a restoring force.
Since $C\neq0$, the beam will be pulled from $k\neq0$ to $k=0$ and acquire acceleration,
which corresponds to an inverse pulling from $x=0$ to $x\neq0$ in the real space.
The direction of the restoring force felt by the incident beam is reversed when it goes across the axis,
but the velocity of the beam does not change its direction, so it will drive the beam from $k=0$ to $k\neq0$, that is, from $x\neq0$ to $x=0$ in the real space.
As it happens, the restoring force in
the $k$ space is constant, thus in the real space, according to the
FSE, it is not harmonic but impulsive, with sudden momentum changes.
As a result, the beam as it propagates is reflected at the places where $|x|$ reaches maximum.

\begin{figure}[htbp]
  \centering
  \includegraphics[width=0.5\columnwidth]{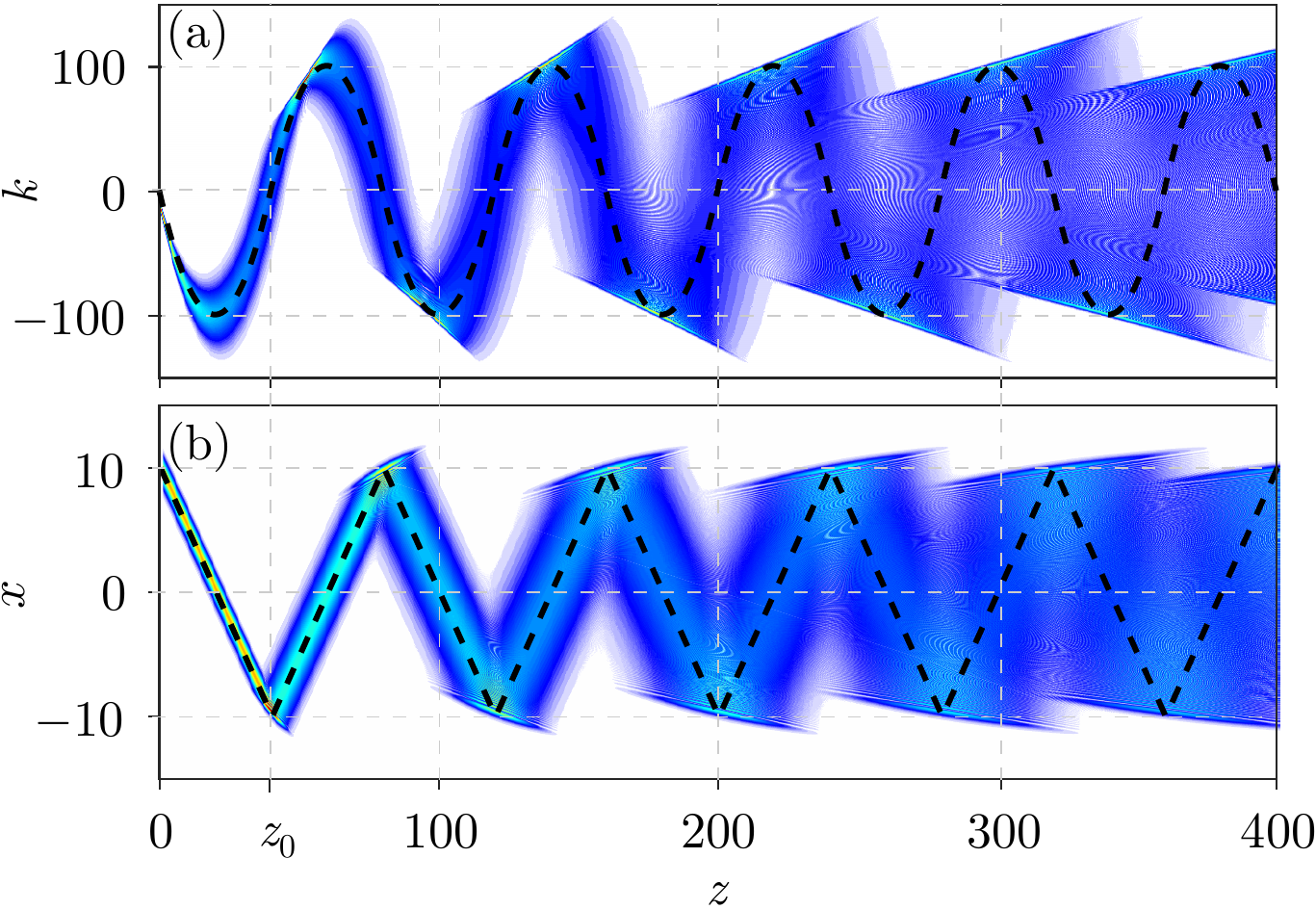}\\
  \caption{(Color online) Propagation of a displaced Gaussian, with $C=0$ and $x_0=10$. Other parameters are as in Fig. \ref{58fig2}.}
  \label{58fig3}
\end{figure}

Even though the beam is well localized in the harmonic potential, it
still broadens during propagation, which can be recognized in Fig.
\ref{58fig2}(b). Since the beam follows a zigzag trajectory during
propagation, the ``incident'' and ``reflected'' beams interfere with
each other at the reflection points. This causes fringes or a
modulation of the beam intensity at the turning points. Such
turnings also increase the spectral width of the beam.

In Fig. \ref{58fig3}, we display the propagation of a displaced
beam, with $C=0$ and $x_0=10$. Again, the analytical trajectories
agree well with the numerical simulations, both in the real and $k$
spaces, at short distances. However, at long distances an irregular
behavior appears, because more and more spatial frequencies are
generated in the $k$ space. This instability should not be confused
with the instability in nonlinear optics. It comes from the fact
that the spatial evolution in FSE involves a path integral
over all interfering rays -- or, in the $k$ space,
it represents a nonlocal operation that involves a continuum of wavenumbers.
Here, Fourier optics is evoked to realize the fractional Laplacian, which is represented through a kernel and leads to a spectral
broadening of the input beam in the $k$ space \cite{longhi.ol.36.2883.2015}. Thus, unlike the usual
quantum mechanics, an input chirped displaced Gaussian broadens
in both $k$ and $x$ spaces, until becoming a featureless broad beam.
It remains bounded, with the power unchanged, oscillating back and forth as a broad pulse.

\begin{figure}[htbp]
  \centering
  \includegraphics[width=0.5\columnwidth]{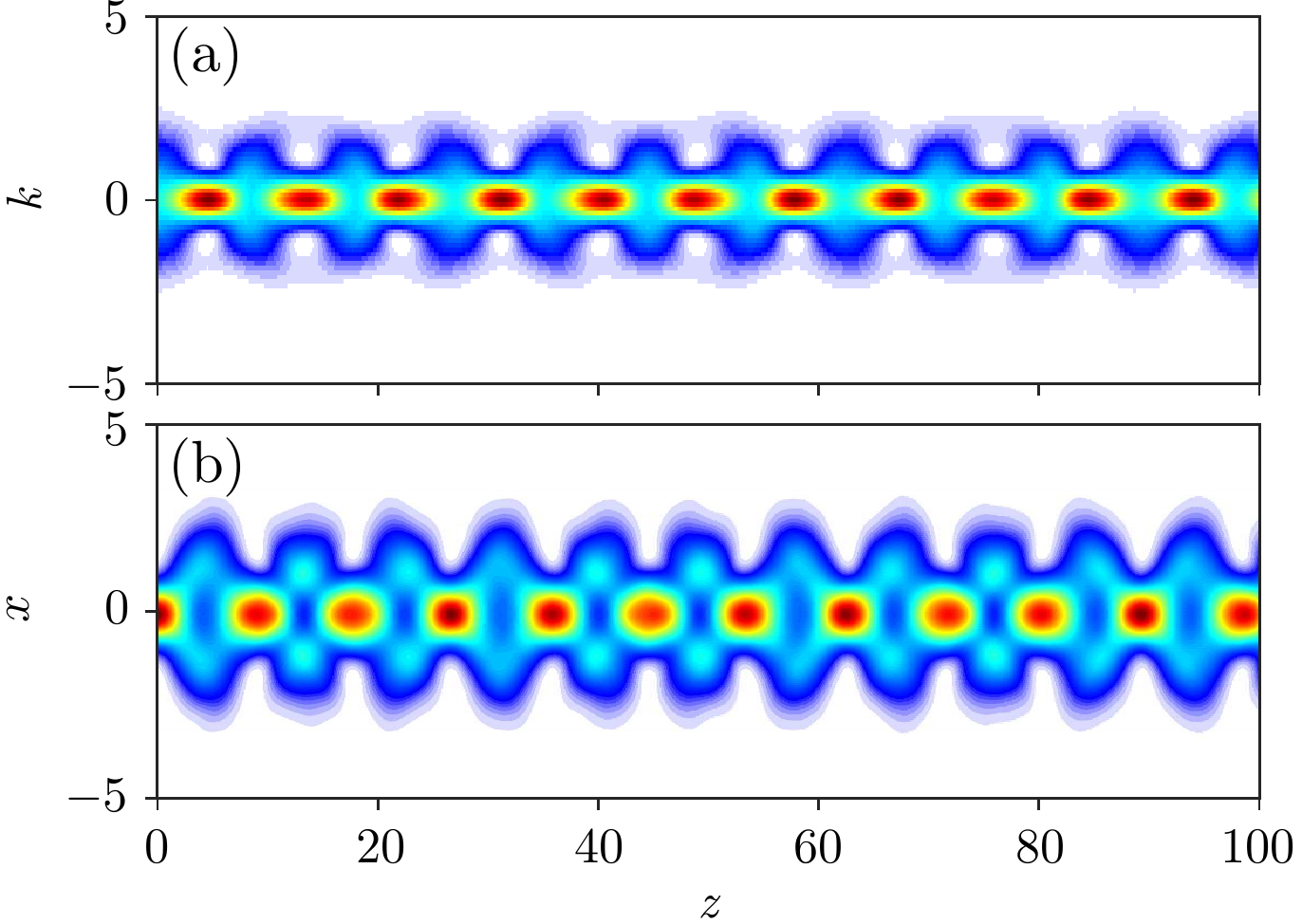}\\
  \caption{(Color online) Same as Fig. \ref{58fig2}, but with $C=0$ and $\beta=0.5$.}
  \label{58fig4}
\end{figure}

As is visible in Eq. (\ref{58eq10}), the period is $\mathcal{Z}=0$ when $x_0=C=0$,
so that the beam will be localized along the $x=0$ and $k=0$ axes during propagation \cite{zhang.jo.17.045606.2015}.
Numerical simulations in the $k$ and $x$ spaces confirm that (Figs. \ref{58fig4}(a) and \ref{58fig4}(b)).
It is clear that the beam exhibits breather-like behavior in both spaces,
which is also a consequence of the nonlocal interference during anharmonic oscillation.

\begin{figure}[htbp]
    \centering
    \includegraphics[width=0.5\columnwidth]{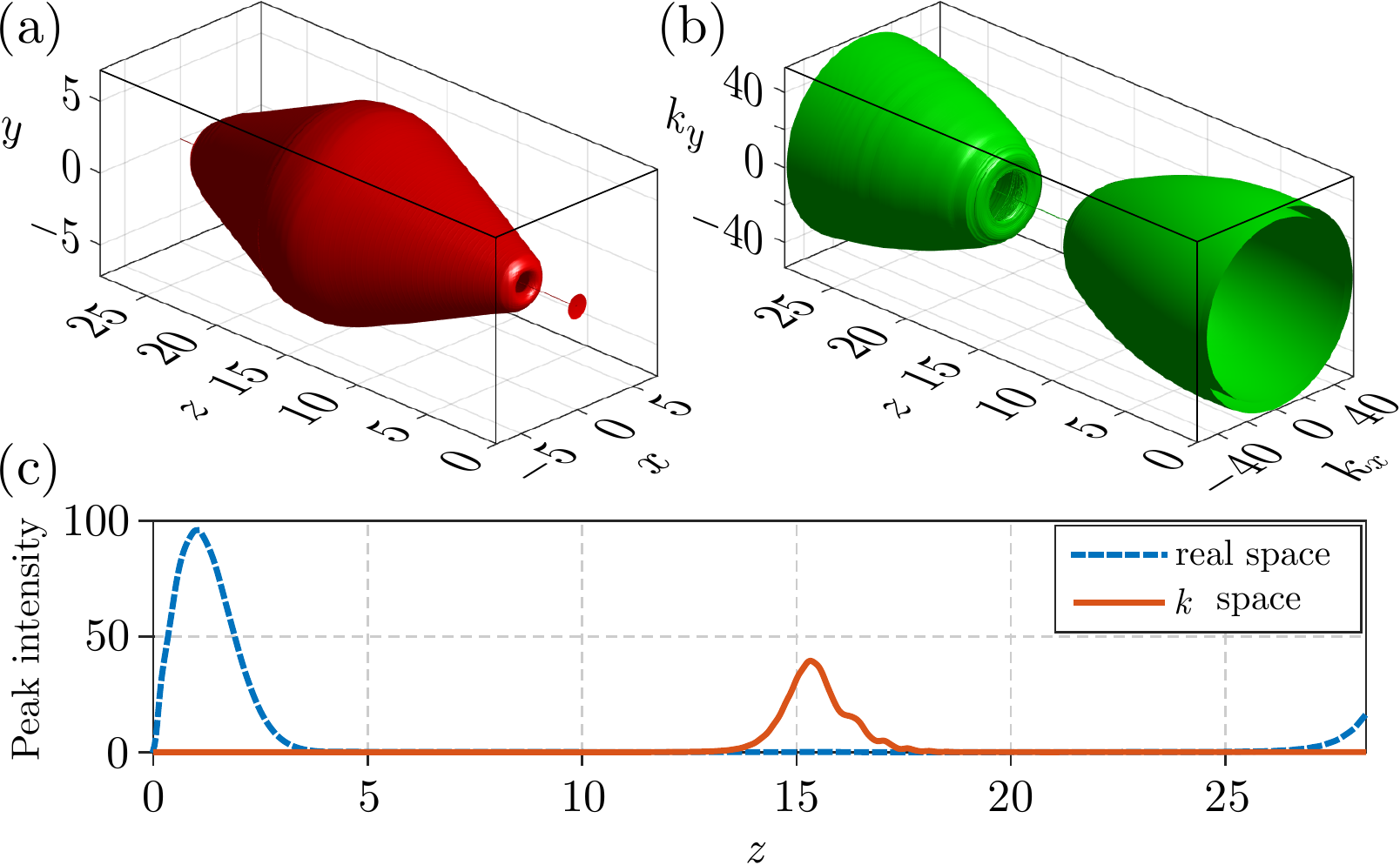}\\
    \caption{(Color online) Isosurface plot of the propagation of a two-dimensional Gaussian beam in (a) real space and (b) $k$ space.
        (c) Peak intensity of the beam versus the propagation distance.
        Parameters are the same as in Fig. \ref{58fig2}.}
    \label{58fig5}
\end{figure}

Now, we turn to the 2D case. The guiding equation is rewritten as
\begin{equation}
i\frac{\partial \psi}{\partial z} = \left[ \frac{1}{2} \left(-\frac{\partial^2}{\partial x^2}-\frac{\partial^2}{\partial y^2}\right)^{\alpha/2} + \frac{\beta^2}{2} \left(x^2+y^2\right) \right] \psi.
\end{equation}
We assume that the incident beam is
$
\psi(x,y)=\exp(-\sigma r^2) \exp(-iCr),
$
with $r^2=(x-r_0)^2+(y-r_0)^2$ and $r_0$ the transverse displacement.
Clearly, the potential in the real space has a parabolic surface profile, and in the $k$ space it will be funnel-like.
Since the incident Gaussian beam is circularly symmetric,
the period for the 2D case is
\begin{equation}
  \mathcal{Z}=4\sqrt{r_0^2 \mp \frac{C}{\beta^2}}.
\end{equation}
The corresponding numerical simulations are shown in Figs. \ref{58fig5}(a) and \ref{58fig5}(b) for the real space and the $k$ space.
The propagation is also indicated in the left and right panels in the movie in Supplementary Material
\footnote{See Supplemental Material at http://link.aps.org/supplemental/ for the animated
version corresponding to Fig. \ref{58fig5}.}.

Since the 2D case is more easily becoming unstable and irregular
than the 1D case, we only display the propagation over one period.
In the real space, the beam evolves to a funnel-like profile, while
in the $k$ space, it propagates along a parabolic surface. The rules
of analysis are similar to those for the 1D case. The interesting
new feature is that now a filamented structure forms during
propagation, both in the real and $k$ spaces, which means that at
places the beam width becomes very small and the peak intensity very
large, as displayed in the movie \cite{Note1}. The physical reason
is that the potential enforces the refractive index change at the
turning points to be kink-like in 1D and cone-like in 2D. So, in the
2D case, the intensity of the propagating beam will focus along the
$x=y=0$ line, which will form a filamented structure, as shown in
Fig. \ref{58fig5}(a). In the Fourier domain, as shown in Fig.
\ref{58fig5}(b), the process unfolds opposite to the real space. The
beam spread is big at the focus in the real space, and small where
the beam is wide. To see the big change in the peak intensity of
the beam during propagation, we depict the intensity in Fig.
\ref{58fig5}(c).

We believe our research not only deepens understanding of the FSE,
but also may have potential applications in the fabrication of light
modulators, in signal processing, and other areas connected with
harmonic potential. In addition, this research may open a way to
investigate other fractional oscillation processes (e.g., fractional
Langevin equation and fractional integral-differential operators
\cite{eab.pa.371.303.2006,*lim.pla.355.87.2006}), and even dynamics
of nonlinear FSE \cite{uchaikin.book.2013}.

This work is supported by National Basic Research Program of China (2012CB921804),
National Natural Science Foundation of China (61308015, 11474228),
Key Scientific and Technological Innovation Team of Shaanxi Province (2014KCT-10),
Natural Science Foundation of Shaanxi Province (2014JQ8341),
and Qatar National Research Fund  (NPRP 6-021-1-005).

%

\end{document}